\documentstyle[preprint,aps]{revtex}

\begin{document}
\draft
%
%
%
\preprint{YCTP P-20-93}
\title{
The Mixmaster Spacetime,  Geroch's Transformation and Constants of
Motion}
\author{
Boro Grubi\v{s}i\'{c} and Vincent Moncrief}
\address{Department of Physics, Yale University,
217 Prospect St., New Haven, CT 06511
}
\date{\today}
\maketitle
\begin{abstract}
We show that for $U(1)$-symmetric spacetimes on $S^3 \times R$
a constant of motion associated with the well known Geroch
transformation, a functional $K[h_{ij},\pi^{ij}]$, quadratic in
gravitational momenta, is  strictly positive in an open
subset of the set of all $U(1)$-symmetric initial data,
and therefore not weakly zero.
The Mixmaster initial data appear to be on the boundary of that
set.  We calculate the constant of motion
perturbatively for the Mixmaster spacetime and find
it to be proportional to the minisuperspace Hamiltonian to the
first order in the Misner anisotropy variables, i.e. weakly zero.
Assuming that $K$ is exactly zero for the Mixmaster spacetime,
we show that Geroch's transformation,
when applied to the Mixmaster spacetime, gives
a new \mbox{$U(1)$-symmetric} solution of the
vacuum Einstein
equations, globally defined on
\mbox{$S^2 \times S^1 \times R$},
which is  non-homogeneous and presumably
exhibits Mixmaster-like complicated dynamical behavior.
\end{abstract}
\pacs{PACS number(s): 04.20.Cv, 04.20.Jb}
\narrowtext

\section{Introduction}
The homogeneous cosmological models, apart from their possible
cosmological relevance, provide us with a rich theoretical laboratory
in which, it is hoped, many of the problems of General Relativity
can be reduced to
a manageable level of complexity. One of the longstanding
problems of Classical Relativity that benefited from such an approach
is the problem of behavior of the solutions of
Einstein's equations close to the ``singularity", i.e.,
close to the boundary of the maximal globally hyperbolic development;
a problem which is  relevant to the issue
of Strong Cosmic Censorship and possibly to the quantization of
gravity program. Due to the complex nonlinear nature of
the Einstein partial differential equations,
an exact and full description of  the asymptotic
behavior seems to be unattainable, at least so in the forseeable future.
Restricted  to the spatially homogeneous cosmological models, however,
the Einstein field equations, in the natural symmetry-adapted foliation,
reduce to a system of ordinary differential equations which is much
easier to analyze, and had been extensively analyzed in the past
(see \cite{ryan-shepley}).
Among the homogeneous models the most important special cases for our
theoretical laboratory are the Kasner and Mixmaster models. The Kasner
solution is explicitly known and is a very simple
(generically) curvature singular
solution. The Mixmaster model, despite several decades of
intense effort, has not yet been explicitly solved (we do not have as
detailed knowledge of its properties as we would want to), but an
approximate description for the approach to the singularity has
been found in terms of a discrete sequence of Kasner solutions
\cite{bkl71}. Due to the
stochastic properties of the associated discrete
mapping it is hard to estimate the quality of the approximation
analytically, but
numerical simulations  indicate that the discrete sequence
approximates the  exact solution the better the closer we come to the
singularity \cite{berger93}.

The significance  of the Mixmaster model for the problem of the
asymptotic behavior of solutions of Einstein's equations stems
largely from the
long series of
papers by Belinski, Khalatnikov and Lifshitz \cite{bkl82}, hereafter
BKL.
 They tried to describe the
approach to the singularity for {\em non-symmetric} spacetimes as a sequence
of pointwise Mixmaster-like transitions between Kasner epochs.
A satisfactory geometrical formulation of their method is still
lacking, and it  is still controversial whether
their method can give any information about the global structure of
the singularity \cite{barrow-tipler}. Nevertheless, it indicates that even
locally the dynamical behavior close to the singularity can be
extremely complex; at least as complex as in the Mixmaster model
(in fact as complex as in non-diagonal Bianchi IX and Bianchi VIII).
Even though the asymptotic dynamics {\em}can be extremely complex
for generic
solutions, there exists a class of solutions  whose asymptotic
behavior is fairly simple: the
asymptotic dynamics simplifies significantly for spacetimes for
which, in a suitable foliation, the spatial derivative terms can be
neglected asymptotically.
Since the thus truncated Einstein's equations
are exactly solvable we can extract all the asymptotic properties we
need from the solution of the truncated equations,
the Generalized Kasner Solution; a solution
which evolves  pointwise
like the Kasner solution.
The corresponding approximation method,
usually called the
Velocity-Dominated Approximation (VDA) \cite{els72,im90} or
Strong Coupling Expansion has been shown to be applicable to
a large class of non-homogeneous spacetimes including Gowdy
spacetimes
\cite{grubisic-moncrief}.
Although the class of spacetimes for which the VDA is valid is
infinite dimensional, it does not include some of the homogeneous
models. Namely, the necessary
conditions for the applicability of the VDA
are not satisfied for the Bianchi VIII and IX spacetimes and, as
BKL showed, are not satisfied for a generic
spacetime as well.

In order to test whether the Mixmaster solution (diagonal Bianchi IX),
or the more general
non-diagonal Bianchi VIII and IX, as claimed by BKL,
are in some sense good asymptotic,
pointwise approximations for the generic spacetimes, as the Kasner
solution is for the velocity-dominated spacetimes, it is important to
have some non-homogeneous solutions that are close to the  Mixmaster
in the gravitational phase space. If this is the case it is also
important to have as detailed knowledge of  the complex Mixmaster
dynamics as possible.

To that end in this paper we
use the structure associated with the well known Geroch
solution generating technique
for spacetimes with one Killing field \cite{geroch71}.
Most importantly,
with Geroch's transformation there is associated a new non-trivial
constant of motion, a
spacetime observable which, we hope,
could help us in the analysis of the asymptotic
behavior of spacetimes.
We  treat the Mixmaster spacetime as a special case of
$U(1)$-symmetric spacetimes
by using only one of the three Mixmaster Killing fields---one
that generates a one-dimensional isometry group  $U(1)$.
The action of that $U(1)$ group on any Mixmaster homogeneous spacelike
slice $\Sigma$,
$\Sigma \sim S^3$, induces a Hopf fibration of
$\Sigma$, i.e., makes it into the $n=1$, $U(1)$ principal fiber bundle
over $S^2$.
We then calculate perturbatively
the new constant of motion for the Mixmaster spacetime
which, to the first order in the
Misner anisotropy variables, turns out to be proportional to the
minisuperspace Hamiltonian.
This, barring an unlikely coincidence, indicates that the
constant of motion is proportional to the minisuperspace
Hamiltonian.
If that is true, we can, using Geroch's transformation, obtain
a new {\em globally} (in space) defined, non-homogeneous
solution on $S^2 \times S^1$, with presumably
Mixmaster-like complex asymptotic behavior.

However, when calculated for the
Taub-Nut special case, a spacetime that has 4 Killing fields,
the constant of motion associated with the fourth Killing field,
which does {\em not exist} in the general Mixmaster case,
is strictly positive and therefore not proportional to the
Hamiltonian constraint (which is zero for the solutions of the Einstein
equations). It is worth emphasizing that there is no contradiction in
the Mixmaster constant of motion being zero and the Taub-NUT
constant being
positive,
since the above mentioned Taub-NUT constant of motion is not a
restriction of the Mixmaster constant of motion to the Taub-NUT
symmetry-class because a different
Killing field is used for the Taub-NUT case.

The outline of the paper is as follows: in section 2 we
review the partial reduction of the  Einstein equations in the
ADM hamiltonian formulation for the $U(1)$-symmetric spacetimes
as well as the conditions for global applicability of Geroch's
transformation. In section 3 we describe the Mixmaster spacetime
as a $U(1)$-symmetric spacetime. In section 4
we apply Geroch's transformation to the
Mixmaster and Taub-NUT spacetimes and find their respective constants
of motion. In section 4 we give some concluding remarks.

\section{$U(1)$-symmetric Spacetimes and Geroch's Transformation}

As shown by Geroch, the vacuum Einstein equations for the class of
four dimensional
spacetimes with at least one Killing field can be reduced,
{\em locally},
to the three dimensional Einstein equations coupled to a harmonic map,
with the two-dimensional
hyperbolic space as the target manifold for the harmonic map. The
isometry group, $SL(2,R)$, of the target space becomes a symmetry of
the reduced equations, i.e., maps solutions of the vacuum Einstein equations
into new locally defined
vacuum solutions with one Killing field.

It was hoped that, with some control over the global
structure of the spacetime, the
transformation might be applicable {\em globally}.
Indeed, for the globally hyperbolic
spacetimes foliated by spacelike, compact, connected
and orientable 3-surfaces (Cauchy surfaces) $\Sigma$,
invariant with respect to the action of the
isometry group, the conditions for generation of new globally (in
space) defined
vacuum solutions were found by Moncrief and Cameron
\cite{moncrief86,cameron-moncrief}.
They showed that the
Einstein equations for U(1)-symmetric spacetimes on $U(1)$
principal fiber bundles (circle bundles)
$\Sigma \times R \rightarrow \tilde \Sigma \times R$---where the
base manifold
$\tilde \Sigma\sim \Sigma/U(1)$ is a compact,
connected  and orientable two-dimensional manifold---can be reduced to a
$2+1$ Einstein-Harmonic map system from $\tilde \Sigma \times R$
to the Poincar\'e
half-plane (hyperbolic space) as the target space,
provided one integrality condition is satisfied
\cite{moncrief86,cameron-moncrief}. For each base manifold $\tilde \Sigma$
there is a countable infinity of isomorphism classes of principal
$U(1)$-bundles $B_n$ over $\tilde \Sigma$,
and the integrality condition depends only on the integer $n$
characterizing the isomorphism class $B_n$ of the Cauchy surface
$\Sigma$. For $\tilde \Sigma=S^2$
all classes were
explicitly constructed by Quiro\'s et. al. \cite{quiros} by imposing
suitable identifications on the $S^3$.
The integer $n$ characterizing the classes
is equal to both the winding number and the Chern number of the
bundle; the
trivial bundle $S^2 \times S^1 \sim B_0$, i.e., $n=0$ and
$S^3 \sim B_1$, i.e., $n=1$, for example.

The restriction to the $U(1)$ isometry group, i.e., the restriction
that the group orbits be closed, is not a restriction for
the spacetimes with at least one dimensional isometry groups.
The isometry group of a cosmological spacetime
must be a compact Lie group---because it is also the
isometry group of the
foliation surfaces $\Sigma$, which are compact Riemannian manifolds---and
every compact Lie group has a $U(1)$ subgroup.
The only real restriction is that there exists a
$U(1)$ subgroup of the isometry group whose action on $\Sigma$
makes the foliation surfaces $\Sigma$
into $U(1)$ principal fiber bundles with compact
and orientable base manifold $\tilde\Sigma$, i.e., that there are no
fixed points for the action and no orbits twisting around each other.

We shall now, mostly following \cite{moncrief86},
give a brief review of Geroch's formalism.
In order to
express the conditions for the global applicability of Geroch's
transformation as  conditions on the gravitational initial data,
we shall
use the usual ADM hamiltonian formulation adapted to the class of
spacetimes with one spacelike Killing field. Depending on convenience,
we shall use index-free or typical-component (abstract-index) notation
for tensor fields. The lower case greek letter ($\mu$, $\nu\cdots$)
indices shall denote components in an arbitrary basis and will run
from 0 to 3. When dealing with coordinates (or tetrads) specially
adapted to a family of spacelike surfaces we shall use the lower case
latin indices, beginning with $i$, for the range from 1 to 3,
and  call such components the spacelike components. Likewise, when
using further specialized coordinate systems (or tetrads), specially
adapted to the group action, we shall use the lower case latin
indices, beginning with $a$, for the range from 1 to 2.

Let the $U(1)$ isometry group be parametrized by the usual angle
parameter $\alpha \in [0,2\pi)$ and let the Killing field $\xi$
be just the derivative with respect to the group parameter
$\alpha$, i.e.,
$\xi[f(p)] = \frac{d}{d \alpha} f( e^{i \alpha}p),$
where $e^{i \alpha} p$ denotes the action of the $U(1)$ group on the
point $p$.
By definition,
the spacetime can be foliated by a family of spacelike hypersurfaces
$\Sigma_t$ invariant
with respect to the action of the isometry group, i.e., a family whose
defining global time function $t$
has zero Lie derivative with respect to the
Killing vector field $\xi$
generating the isometry group action. The Killing
field is then necessarily tangent to the foliation.

As usual, using the unit normal vector field $n^\mu$ to $\Sigma$
and the spacetime metric $g_{\mu\nu}$ one can define a new tensor
field, the three-metric
\begin{equation}
h_{\mu\nu}= g_{\mu\nu} + g_{\mu\lambda} \, g_{\nu\sigma} \;
 n^\lambda n^\sigma,
\end{equation}
whose restriction to the vectors tangent to $\Sigma$ generates a
Riemannian metric on $\Sigma$.
In a coordinate
system generated by extending  arbitrary coordinates on $\Sigma$
using the flow of the normal vector $n^\mu$, only spacelike components
$h_{ij}$ are non-zero, and
both the three-metric $h_{\mu\nu}$ and the foliation unit
normal vector field $n^\mu$ are group invariant.

The three-metric
$h_{\mu\nu}$ can be further decomposed as follows:
\begin{equation}
h_{\mu\nu}= \lambda^{-1} \tilde g_{\mu\nu} + \lambda \;\eta_\mu
\eta_\nu,
\end{equation}
using the norm $\lambda$ of the Killing field $\xi^\mu$,
\begin{equation}
\lambda= g_{\mu\nu} \xi^\mu \xi^\nu = h_{\mu\nu} \xi^\mu \xi^\nu,
\end{equation}
and a one-form $\eta_\mu$ defined as
\begin{equation}
\eta_\mu= \lambda^{-1} g_{\mu\nu} \xi^\nu.
\end{equation}
This decomposition is $\xi$-invariant  and induces a new Riemannian
two-dimensional metric $\tilde g_{\mu\nu}$ in the subspace of vectors
tangent to $\Sigma$ and orthogonal to $\xi$.
As a result one gets a $\xi$-invariant decomposition of the metric
$g_{\mu\nu}
\rightarrow (n_\mu,\tilde g_{\mu\nu},\eta_\mu,\lambda)$, and
it is easy to see that
\begin{eqnarray}
\xi^\mu n_\mu&=&0, \\
\xi^\mu \tilde g_{\mu\nu}&=&0, \\
\xi^\mu \eta_\mu&=&1 \label{xi-eta}, \\
\xi^\mu (d\eta)_{\mu\nu}&=&0.
\end{eqnarray}
The last equation being a consequence of (\ref{xi-eta}) and
${\cal L}_{\xi} \eta=0$.

The bundle projection pullback $^\ast\pi$ generates a
one to one correspondence between, on the one side, the
$\xi$-invariant covariant
tensor fields on $\Sigma \sim B_n$
all of whose contractions
with $\xi$ vanish, and on the other, the covariant tensor fields on
the base $\tilde \Sigma$ \cite{geroch72},
We can, therefore, treat the fields   $\lambda$ and $\tilde g_{ab}$
as tensor fields on $\tilde \Sigma$ (and $n_\mu$ as a field on
$\tilde \Sigma \times R$). The one-form $\eta$, however, does not have
vanishing contraction with $\xi$ and cannot be treated as a field on
$\tilde\Sigma$, which
imposes the only global restriction to the reduction of dynamics
from $\Sigma$ to $\tilde\Sigma$. Even though $\eta$ does not have
vanishing contraction with $\xi$, its exterior derivative (the only
form in which $\eta$ will enter the equations of motion) does, and
as shown in \cite{cameron},
for a closed
two-form $\Phi$ on $\tilde \Sigma$
there will exist a one-form $\eta$ on $\Sigma \sim B_n$,
satisfying $\xi^\mu \eta_\mu=1$ and
$d\eta= {^\ast\pi(\Phi)}$,  if and only if
\begin{equation}
\int_{\tilde \Sigma} \Phi=2 \pi n. \label{integrality1}
\end{equation}

To effectively use the symmetry of the problem in concrete
calculations we shall be doing in this paper,
we need an
isometry-adapted atlas on the spacetime manifold.
Using an atlas on the base manifold $\tilde \Sigma$,
with coordinates in a representative chart labeled by
$\{x^a \}$, $a=1,2$  one can construct, with the help
of the bundle structure on $\Sigma$, a $\xi$-invariant atlas on the initial
foliation surface $\Sigma$. Let the coordinates in a representative chart
be labeled $ \{x^i\}=\{x^a,x^3 \}$, $i=1..3$, with the third coordinate
normalized so that
$\partial / \partial x^3 = \xi$.
Now, one can
choose a $\xi$-invariant time-evolution vector field $t^\mu$ and
lift the coordinates from the initial leaf $\Sigma$, along the flow
generated by $t^\mu$, to all other foliation surfaces, which are labeled
by the flow parameter $t$ used as the zeroth coordinate.
Thus, one  has created an
atlas on the spacetime manifold in whose every chart,
$ \{x^\mu\}=\{t,x^a,x^3 \}$, $\mu=0..3$,
the action of the
$U(1)$ group is given by
$e^{i \alpha} (t,x^a,x^3) = (t,x^a,x^3 + \alpha),$
and the bundle projection $\pi$ by
$\pi : (t,x^a,x^3) \mapsto (t,x^a)$.

In any such chart we can define a one-form
\begin{equation}
\beta= \eta - dx^3,
\end{equation}
which carries all important dynamical information contained in $\eta$
and whose exterior derivative, $d\beta=d\eta$, and time derivative,
$\dot \beta \equiv {\cal L}_t \beta= \dot \eta$, are globally defined
and chart independent differential forms.
The two-form $d\beta$ can be identified with the (exact,
globally-defined) two-form $\Phi$
on the base manifold $\tilde \Sigma$ and is,
in an ${x^a}$ chart, expressible as
\begin{equation}
d\beta= d\eta=r \; dx^1 \wedge dx^2, \label{deta}
\end{equation}
where $r$ is a scalar density whose value in the same chart,
\begin{equation}
r= \epsilon^{ab} \beta_{a,b}, \label{r}
\end{equation}
can be calculated using the contravariant
antisymmetric tensor (density) $\epsilon^{ab}=-\epsilon^{ba}$,
$\epsilon^{12}=1$,
associated with the chart.

In the new $U(1)$-adapted variables the isometry of the spacetime
metric,
\begin{eqnarray}
ds^2=g_{\mu\nu}\;dx^\mu dx^\nu &=&
-N^2 dt^2 + h_{ij} ( dx^i + N^i dt) ( dx^j + N^j dt)  \nonumber
\\
  &=&\lambda^{-1} \left[
- \tilde N dt^2 + \tilde g_{ab} (dx^a + \tilde N^a dt)
(dx^b + \tilde N^b dt)
\right] \nonumber \\
 &&+ \lambda \left[
 dx^3 + \beta_a dx^a + \tilde N^a \beta_a dt \right]^2,
\label{admform}
\end{eqnarray}
is equivalent to all the
component functions being independent of $x^3$,
as the coordinate one-forms $\{dt,dx^i\}$
are already $\xi$-invariant.

Using the $U(1)$ isometry, the ADM action $I$ on the $\Sigma\times R$,
\begin{equation}
I= \int dt \int_\Sigma dx^3 \left
\{ { \pi^{ij} h_{ij,t} - N {\cal H} - N^i {\cal H}_i}
\right\},
\end{equation}
with
\begin{eqnarray}
{\cal H} &=& h^{-1/2}
 \left[ \pi^{ij} \pi_{ij} -
1/2 (\pi^i_i)^2  \right] - h^{1/2} R(h), \\
{\cal H}_i &=& -2 h^{1/2} \nabla_j ( h^{-1/2} \pi^j_i ),
\end{eqnarray}
reduces to an equivalent action $\tilde{I}$ on
$\tilde\Sigma \times R$,
\begin{equation}
\tilde I = 2 \pi \int dt \int_{\tilde\Sigma} dx^2 \left\{
\tilde \pi^{ab} \tilde g_{ab,t} + p \lambda_{,t}+ e^a \beta_{a,t}
- \tilde N \tilde {\cal H} - \tilde N^a \tilde{\cal H}_a  + \beta_0
  e^a_{,a} \right\},
\end{equation}
where
\begin{eqnarray}
\tilde {\cal H}&=&
 \lambda^{-1/2} {\cal H} =  \tilde g^{-1/2}
 \left[
 \tilde \pi^{ab} \tilde \pi_{ab} - \left( \tilde \pi^a_a \right)^2
+ \frac{1}{2} \lambda^2 p^2 + \frac{1}{2 \lambda} \tilde g_{ab}
e^a e^b \right] \nonumber \\
&+& \tilde g^{1/2} \left[
- R(\tilde g)
+ \frac{1}{2 \lambda} \tilde g^{ab} \lambda_{,a} \lambda_{,b}
+ \frac{\lambda^2}{4} \tilde g^{ac} \tilde g^{bd} (d\beta)_{ab}
(d\beta)_{dc} \right], \\
\tilde{\cal H}_a &=& {\cal H}_a =
- 2 \tilde g^{1/2} \tilde \nabla_b  \left(
\tilde g^{-1/2} \tilde \pi^b_a  \right)
+ p \lambda_{,a} + e^b (d\beta)_{ab}, \\
\tilde N &=& \lambda^{1/2} N,\\
\tilde N^a &=& N^a  , \\
\beta_0 &=& \eta_i N^i, \\
\tilde \pi^{ab} &=& \lambda^{-1} \pi^{ab}, \\
e^a &=&2  \lambda \eta_i \pi^{ia} , \label{e} \\
p &=&  2 \pi^{ij} \eta_i \eta_j - \lambda^{-1} \pi^{ij} h_{ij} .
\end{eqnarray}

Variation of the action $\tilde I$ with respect to $\beta_0$ gives a
constraint:
\begin{equation}
e^a_{,a}=0 \Leftrightarrow {\rm div\;} \overline{e}=0
\Leftrightarrow  d(*\overline{e})=0;
\end{equation}
where we have used $\overline{e}= \tilde g^{-1/2} \; e$ to denote
vector field associated to the vector density $e$, and * to denote the
Hodge star operator of the Riemannian
metric $\tilde g$ on $\tilde \Sigma$.
This constraint can be easily solved since,
according to the Hodge theorem, every closed form on $\tilde \Sigma$,
and in our case this is the form $*\overline{e}$,
can be uniquely decomposed  to a sum of one exact and one harmonic
form. Explicitly written in an arbitrary chart on $\tilde \Sigma$,
the decomposition gives
\begin{equation}
e^b \epsilon_{ba}  = \omega_{,a} + h_a, \label{hodge}
\end{equation}
with $\omega$ the scalar field (defined up to a constant)
and $h_a$ the components of the unique harmonic form,
both defined globally on $\tilde \Sigma$. By a slight abuse of
notation, we have used $\epsilon_{ab}=-\epsilon_{ba}$,
$\epsilon_{12}=1$, to denote the
covariant antisymmetric tensor (density) associated with the chart.
{}From  the equations of motion obtained by varying
the action with respect to $\beta_a$, we can obtain
\begin{equation}
\dot \omega_{,a} + \dot h_a = \left[
\tilde N \tilde g^{-1/2} \lambda^2 r + \tilde N^b (\omega_{,b} + h_b)
\right]_{,a}, \label{varbeta}
\end{equation}
which, because of the uniqueness of Hodge decomposition and the fact that
the one form on the right-hand side is exact,
forces the harmonic form $\dot h$ to be zero, i.e., $h$ to be time
independent. Hereafter in this paper we shall be interested only in the special
case of $\tilde \Sigma \sim S^2$ which has no non-zero harmonic
one-forms,
and shall therefore put $h=0$.

With the decomposition of $e^a$ included into the
action $\tilde I$, the one-form
$\beta_a$ appears only in the form $\epsilon^{ab} \beta_{a,b}=r$
and the action reduces to an equivalent
2+1 Einstein-Harmonic action
\begin{equation}
\tilde J = 2 \pi \int dt \int_{\tilde\Sigma} dx^2 \left\{
\tilde \pi^{ab} \tilde g_{ab,t}  + p \lambda_{,t}+ r \omega_{,t}
- \tilde N \tilde {\cal H} - \tilde N^a \tilde{\cal H}_a  \right\},
\end{equation}
where
\begin{eqnarray}
\tilde {\cal H}&=&  \tilde g^{-1/2}
 \left[
 \tilde \pi^{ab} \tilde \pi_{ab} - \left( \tilde \pi^a_a \right)^2
+ \frac{1}{2}
G^{AB} P_A P_B
 \right] \nonumber \\
&+& \tilde g^{1/2} \left[
- R(\tilde g)
+ \frac{1}{2}
G_{AB} X^A_{,a} X^B_{,b} \tilde g^{ab}
 \right], \\
\tilde{\cal H}_a &=&
- 2 \tilde g^{1/2} \tilde \nabla_b  \left(
\tilde g^{-1/2} \tilde \pi^b_a  \right)
+ p \lambda_{,a} + r \omega_{,a}, \\
X^A &=& \{\lambda,\omega\}, \quad P_A = \{p,r\}, \quad A=1,2,
\end{eqnarray}
and
\begin{equation}
G_{AB}= \lambda^{-2} \left[ d\lambda^2 + d\omega^2 \right]_{AB}
\end{equation}
is the metric of constant negative curvature on the Poincar\'{e}
half-plane $\{\lambda >0,\omega\}$, the target space for the harmonic
variables $\lambda$ and $\omega$.
This form of the action makes evident the $SL(2,R)$ symmetry of the
equations of motion,
$SL(2,R)$ being the isometry group of the Poincar\'{e} half-plane,
 and enables us to apply the Geroch transformation to the harmonic
variables $\lambda$ and $\omega$ to generate new solutions of the
vacuum 3+1 Einstein equations from the known ones.

Starting with a $U(1)$-symmetric four-dimensional
solution of the vacuum Einstein equations on
$R \times B_n$ expressed in the form (\ref{admform}), we can calculate the
canonical variables ($\lambda ,\,  r,\,p$) locally from the metric
components and their velocities.
$\lambda$ is the norm of the Killing vector field, $r$ is
given by (\ref{r}) and
\begin{equation}
p= 2 \tilde N^{-1} \tilde g^{1/2} \lambda^{-1}
\left( \dot \lambda - \tilde N^a \lambda_{,a} \right). \label{p}
\end{equation}
Only the  canonical variable $\omega$ is not a local function of the
metric components and their velocities (or momenta), but it can be
calculated by performing a
line integral in the three-dimensional space $R\times \tilde \Sigma$:
\begin{equation}
\omega(t,x^1,x^2)= \int_{t_0}^t \dot \omega (t',x^1_0, x^2_0) \;dt +
 \int_{\Gamma (x_0,x)} \omega_{,a}(t,x') \;dx'^a . \label{intomega}
\end{equation}
In order to do so,
we need the spatial and temporal derivatives of $\omega$.
Without using the equations of motion in the Einstein-Harmonic form
the spatial derivatives can be calculated
from (\ref{e}) and (\ref{hodge}) and the time derivative from
(\ref{varbeta}) giving:
\begin{eqnarray}
\omega_{,a}&=& - \tilde N^{-1} \tilde g^{1/2} \lambda^2
\epsilon_{ab} \;\tilde g^{bc} \dot \beta_c ,\\
\omega_{,t}&=& \tilde N \tilde g^{-1/2} \lambda^2 r - \tilde N^{-1}
\tilde N^a \tilde g^{1/2} \lambda^2
\epsilon_{ab} \;\tilde g^{bc} \dot \beta_c + c(t), \label{dotomega}
\end{eqnarray}
where $c(t)$ is an arbitrary function of time.
This is to be expected,
since $\omega$ is defined only up to a constant on each $\tilde \Sigma$,
i.e., given an $\omega$ a whole class of functions
$\omega + f(t)$ is associated to each
four dimensional spacetime. So, it would appear that there is a gauge
indeterminacy for the evolution of $\omega$. Fortunately, the new evolution
equations completely eliminate that gauge freedom for $\omega$ and,
given initial data for $\lambda, \omega, p$ and $r$, determine a unique
solution.
Varying the action $\tilde J$ with respect to $r$,
we find that the class representative for $\omega$
singled out by the new equations of motion
is the one obtained by putting $c(t)=0$ in (\ref{dotomega}). The class
representative is not completely determined---a constant (in time and
space) can be added to $\omega$---which corresponds to freedom of
choice of the reference point $x_0$
in the initial $\tilde \Sigma$ from which the spatial
line integration in (\ref{intomega}) starts.

The action of an element,
\begin{equation}
\left(
\begin{array}{cc}
a&b\\
c&d
\end{array}
\right), \quad ab-cd=1,
\end{equation}
of $SL(2,R)$ on the canonical variables,  given by:
\begin{eqnarray}
\lambda \rightarrow
\lambda'& =& \frac{\lambda}{
c^2 ( \omega^2 + \lambda^2 ) + 2 c d \omega + d^2 }, \\
\omega \rightarrow
\omega'&=&
\frac{a c (\omega^2 + \lambda^2) + (a d + b c ) \omega + b
d}{c^2 ( \omega^2 + \lambda^2 ) + 2 c d \omega + d^2 }, \\
p \rightarrow
p'&=&
\frac{p\left[c^2(\omega^2-\lambda^2) + 2 c d \omega + d^2
\right] - r \left[2 \lambda (c d + c^2 \omega) \right] }
{c^2 ( \omega^2 + \lambda^2 ) + 2 c d \omega + d^2 }, \\
r \rightarrow
r'&=&
p ( 2 c^2 \lambda \omega + 2 c d \lambda) + r \left[
d^2 + c^2 (\omega^2 -\lambda^2) + 2 c d \omega \right],
\end{eqnarray}
then transforms our old  solution of the 2+1 Einstein-Harmonic
equations
on
$R \times \tilde \Sigma$ into a
new solution of the same equations. To lift that solution  to a new
solution of the 3+1 vacuum Einstein equations  on $R \times B_{n'}$,
all we need to do is use new $\lambda'$ and integrate (\ref{deta})
to find the new one-form
$\eta'$. As mentioned previously, this will be possible if and only
if the integrality condition (\ref{integrality1}), which now becomes
\begin{equation}
\int_{\tilde \Sigma} r' =2\pi n',
\end{equation}
is satisfied at all times.

The above integral is a constant of motion of the Einstein-Harmonic
equations  and the integrality
condition has to be enforced only at one
time. Even better, the symmetry-action
of $SL(2,R)$ results in three conserved quantities, each
associated with one of
the three Killing fields of the Poincar\'{e} half-plane.
These three constants of motion are:
\begin{eqnarray}
A&=& \int_{\tilde\Sigma} (2 \omega r + p), \\
B&=& \int_{\tilde\Sigma} r,\\
C&=& \int_{\tilde\Sigma} \left[ r ( \lambda^2 -\omega^2) - p \omega
\right],
\end{eqnarray}
and represent a momentum map of the $SL(2,R)$ action.

$B$ is not interesting as a constant of motion since its value
does not depend on the initial data; it is just the integrality
condition dictated by the topology of $\Sigma$.  $A$ and
$C$ are not proper constants of motion associated to the
spacetime, either,
because their values depend on the constant added to $\omega$,
which is not observable in the spacetime.
Nevertheless,
the $SL(2,R)$-invariant constant of motion found by Geroch \cite{geroch72},
\begin{equation}
K[h_{ij}, \pi^{ij}]= A^2 + 4 B C. \label{casimir}
\end{equation}
has the desired properties.
It is easy to see that the value of $K$, not only
does not depend on the unobservable constant added to $\omega$, but
remains the same even if we add an arbitrary function of time
to $\omega$, which means that it is a proper constant of motion of the
spacetime, i.e., an observable. This also means that when calculating
$\omega$ to be used for evaluating $K$,
the first term in (\ref{intomega}), which is only
a function of time, can be dropped.
That guarantees that the constant of motion $K$
depends only on the initial data on a single Cauchy surface, i.e.,
that it is a functional of the standard gravitational variables
$(h_{ij}, \pi^{ij})$ that is local in time.

In order to calculate $K[h_{ij},\pi^{ij}]$, given the ADM initial data
$(h_{ij},\pi^{ij})$, one has to express $\lambda$, $p$, $r$ and
$\omega_{,a}$
in terms of the components of the ADM data in a $U(1)$-adapted
coordinate chart:
\begin{eqnarray}
\lambda &=& h_{33}, \\
p&=&2 \lambda^{-2} \pi^{ij} h_{i3} h_{j3} - \lambda^{-1} \pi^{ij}
h_{ij}, \\
r&=& \left( \lambda^{-1} h_{31} \right)_{,2}-
\left( \lambda^{-1} h_{32} \right)_{,1}, \\
\omega_{,a}&=& 2 \epsilon_{ba} \pi^{bi} h_{i3},
\end{eqnarray}
and evaluate $\omega$ by integrating $\omega_{,a}$ along a curve
in the two-sphere. The constants of motion $A$, $B$ and $C$ are then
calculated by evaluating the surface integrals. Loosely speaking,
$K$ is a quadratic functional of the ADM momenta $\pi^{ij}$.

An important property of $K$ is that it controls the global
applicability of the Geroch transformation. As shown by Moncrief
\cite{moncrief87}, any solution on a nontrivial bundle
$B_n$, $n \neq 0$,
can be transformed to a globally defined solution on any other
nontrivial bundle $B_{n'}$, $n' \neq 0$; no restrictions exist.
For a more interesting case
of transforming a solution from a nontrivial bundle to a global
solution on
the trivial bundle there is a restriction, however.  The
transformation can generate a globally defined solution if and only if
$K\geq 0$.

In the next section we shall establish some of the properties of $K$
and try to calculate it  for the Mixmaster spacetime.

\section{Mixmaster Model as a Special Case of the $U(1)$-symmetric
Spacetime}
In a spatially homogeneous spacetime, by definition,
the orbits of the isometry group
(or some subgroup if there are timelike Killing vector fields)
are three-dimensional spacelike hypersurfaces that
foliate the spacetime. If the isometry group has a subgroup that acts
simply transitively on the orbits, the spacetime can be classified into
one of the nine Bianchi classes, according to the Bianchi class of the Lie
algebra of the corresponding simply transitive subgroup. If the
isometry group has no simply transitive subgroup, the isometry group
must be at least four-dimensional and the spacetime
must be locally isometric to the
interior of the Schwarzschild solution. These solutions, found by
Kantowski and Sachs \cite{kantowski-sachs}, if restricted to compact
spatial sections must have the spatial sections diffeomorphic to
$S^2 \times S^1$ and the four dimensional group with transitive action
must be $SU(2) \times U(1)$.

The Mixmaster  cosmological model, by definition,
belongs to the Bianchi IX class,
has $S^3$ spatial topology and $SU(2)$ as the isometry group, and the
spatial metric is diagonal in the left- or right-invariant basis of
the isometry group.

Let us begin with the brief
analysis of the structure of the spatial sections.
The three-sphere, the group manifold of $SU(2)$, can be represented
as a hypersurface in ${\bf C}^2\sim {\bf R}^4$
$ S^3= \left\{ (z_1,z_2), \; |z_1|^2 + |z_2|^2 =1   \right\} $.
Writing $z_1=r_1 e^{\alpha_1}$ and $z_2=r_1 e^{\alpha_2}$, with
$r_i\geq 0$ and $\alpha_i$ from any interval of width $2\pi$,  we see that
the only condition the points lying on $S^3$
must satisfy  is $(r_1)^2 + (r_2)^2 =1$, which guarantees that there
exist
unique $\theta\in [0,\pi]$ such that $r_1=\cos\theta/2$ and
$r_2= \sin\theta/2$. Thus, one can introduce
$\{\theta,\alpha_1,\alpha_2\}$
as coordinates on $S^3$,
and can think of $S^3$ as a singular
one-parameter family
of two-tori $(\alpha_1,\alpha_2)$, $\theta$ being the parameter of the
family. The tori for $\theta=0,\pi$  collapse to $S^1$, making the
coordinate system singular there.

Recall that we
need a coordinate system in which one coordinate vector field would
have closed orbits, i.e., generate a  $U(1)$ group action, and
that group action would have to be such that it gives  $S^3$  a principal
fiber bundle structure over some two-dimensional $\tilde \Sigma$.
The most useful coordinates on $S^3$ , for that  purpose,
are the standard Euler angle coordinates $\{\theta,\phi,\psi\}$,
which can be obtained from $\{\theta,\alpha_1,\alpha_2\}$ coordinates
by a simple coordinate transformation:
\begin{equation}
\phi = \alpha_1 + \alpha_2, \quad \psi= \alpha_1 - \alpha_2.
\end{equation}
The Euler coordinates, just like the $\{\theta,\alpha_1,\alpha_2\}$
have a singularity at
$\theta=0,\pi$ and when we let $\{\theta,\phi,\psi\}$ $\in$ $ \{(0,\pi),
[0,2\pi),[0,4\pi)\}$ or $\{\theta,\phi,\psi\}$ $\in$ $\{(0,\pi),
[0,4\pi),[0,2\pi)\}$
they cover all of the three-sphere (except the two singular circles)
exactly once.  The period of both $\phi$ and $\psi$ coordinates is
$4 \pi$, and if
we let both $\phi$ and $\psi$ extend all the way to $4\pi$, the
three-sphere would be covered twice.

The coordinate vector fields $\partial_\phi$ and
$\partial_\psi$ are globally defined on $S^3$ despite the coordinate
singularities in $\theta=0,\pi$ and they generate two $U(1)$ group
actions, which each yield  a Hopf fibration of $S^3$, i.e., they make
$S^3$ into a principal fiber bundle over $S^2$ with bundle projection for
the $\psi$-bundle, for example, given by:
\begin{equation}
\pi_\psi: S^3 \rightarrow S^2, \quad
(\theta,\phi,\psi)\mapsto (\theta,\phi).
\end{equation}
Since these two bundle structures are equivalent, we shall use
$\psi$-bundle for the Mixmaster spacetime and, taking $\psi/2=x^3$, define
\begin{equation}
\xi=2 \;\partial_\psi,
\end{equation}
to be the Killing field we shall use to apply
Geroch's transformation to the Mixmaster solution.
The coordinate system $\{x^1,x^2,x^3\}=\{\theta,\phi,\psi/2 \}$
is appropriately adapted to the bundle structure, as described in the
previous section.
In a spacetime with several Killing fields it is important to remember
which Killing field was employed for Geroch's transformation,
so we shall denote
the constant of motion (\ref{casimir}) associated with
the Killling vector $2\;\partial_\psi$ by $K_\psi$.

The set of standard, globally-defined (and analytic)
left-invariant vector fields $\hat X_i$,
 which generate a right-action on  $SU(2)$ \cite{miller73},
written here together with their dual one-forms $\hat\omega^i$:
\begin{equation}
\begin{array}{ll}
\hat X_1 &= \cos\psi \;\partial_\theta + \csc\theta \sin\psi
\;\partial_\phi - \cot\theta \sin\psi \;\partial_\psi \\
\hat X_2 &= - \sin\psi \;\partial_\theta + \csc\theta \cos\psi
\;\partial_\phi - \cot\theta \cos\psi \;\partial_\psi \\
\hat X_3 &= \;\partial_\psi \\
\hat \omega^1 &= \cos\psi \;d\theta + \sin\theta \sin\psi \;d\phi \\
\hat \omega^2 &= - \sin\psi \;d\theta  + \sin\theta \cos\psi \;d\phi \\
\hat \omega^3 &= d\psi + \cos\theta \;d\phi
\end{array}
\end{equation}
\begin{equation}
\begin{array}{rl}
\left[ \hat X_i,\hat X_j \right] &= C^k_{\;ij} \;\hat X_k \\
d\hat\omega^k &= -1/2 \;C^k_{\;ij} \;\hat\omega^i \wedge \hat\omega^j \\
C^k_{\;ij} &= \epsilon_{ijk}, \quad \epsilon_{123}=1
\end{array}
\end{equation}
contains $\partial_\psi$,  so, we shall chose them as
the complete set of Killing fields for the Mixmaster spacetime.
We shall also need a right-invariant one-form basis $\hat\sigma^i$
to explicitly
express the isometry of the metric. For completeness we give them with
their dual vector fields $\hat Y_i$ and some of their properties:
\begin{equation}
\begin{array}{ll}
\hat Y_1 &= \cos\phi \;\partial_\theta + \csc\theta \sin\phi
\;\partial_\psi - \cot\theta \sin\phi \;\partial_\phi \\
\hat Y_2 &=  \sin\phi \;\partial_\theta - \csc\theta \cos\phi
\;\partial_\psi + \cot\theta \cos\phi \;\partial_\phi \\
\hat Y_3 &= \partial_\phi \\
\hat \sigma^1 &= \cos\phi \;d\theta + \sin\theta \sin\phi \;d\psi \\
\hat \sigma^2 &= \sin\phi \;d\theta  + \sin\theta \cos\phi \;d\psi \\
\hat \sigma^3 &= d\phi + \cos\theta \;d\psi
\end{array}
\end{equation}
\begin{equation}
\begin{array}{rl}
\left[ \hat Y_i,\hat Y_j \right] &= -C^k_{\;ij} \;\hat Y_k \\
d\hat\sigma^k &= 1/2 \;C^k_{\;ij} \;\hat\sigma^i \wedge \hat\sigma^j \\
\left[ \hat X_i,\hat Y_j \right] &=0, \quad
{\cal L}_{\hat X_i} \hat \sigma_j =0
\end{array}
\end{equation}

The Mixmaster metric, invariant with respect to the right action of
$SU(2)$, can now be written as:
\begin{equation}
ds^2= - N^2 \;dt^2 + A^2  (\hat \sigma^1)^2 +
B^2  (\hat \sigma^2)^2 + C^2  (\hat \sigma^3)^2, \label{ds2}
\end{equation}
where $N$, $A$, $B$ and $C$ are  functions of time only, and when
expressed in the usual Misner anisotropy variables $\Omega$,
$\beta_+ $ and $\beta_- $,
\begin{equation}
\begin{array}{ll}
A &= e^{-\Omega + \beta_+ + \sqrt{3 }\beta_- }, \\
B &= e^{-\Omega + \beta_+ - \sqrt{3 }\beta_- }, \\
C &= e^{-\Omega - 2 \beta_+ } .
\end{array}
\end{equation}

\section{The Constants of Motion and Transformed Spacetimes}

As the  first step in calculating the constants of motion for the
Mixmaster spacetime, we need to
calculate the harmonic variables.
The function $\lambda$
and one-form $\beta$ can be found
simply by expanding the metric (\ref{ds2}). The
functions $r$ and  $p$ can be straightforwardly
calculated using (\ref{r}) and
(\ref{p}), which requires only differentiation. The function $\omega$, on
the other hand, is hard to calculate since that involves a line
integration in the $(\theta$,$\phi)$ surface with
\begin{equation}
\lambda = A^2 \sin^2 \phi \sin^2 \theta + B^2 \cos^2 \phi \sin^2
\theta + C^2 \cos^2\theta
\end{equation}
in the denominator. We have not been able to evaluate $\omega$ in
closed form, which prevented us from calculating the constant of motion
$K_\psi$ exactly. But, even if one succeeded in putting $\omega$ into a
complicated closed form
it would probably be of little practical importance, since we
would still have to evaluate the surface integrals, which seem
extremely difficult.

However, the integrals are simple enough when $A=B=C$, i.e., when
the anisotropy variables
$\beta_+ = \beta_- =0$, which allowed us to calculate $K_\psi$ perturbatively
around the isotropic solution,
to the first order in the anisotropy variables.
This calculation is tedious and rather
unilluminating, so we shall state only the main result:

{\em
To the first order in the anisotropy variables $(\beta_+, \beta_-)$,
the constant of motion
$K_\psi$ for the Mixmaster spacetime
is proportional to the minisuperspace Hamiltonian,
}
\begin{eqnarray}
H&=&1/12 \;N e^{3\Omega} \left[ - p_{\Omega}^2 + p_+^2 + p_-^2 +
e^{-4 \Omega} V(\beta_+,\beta_-) \right], \\
&& V(\beta_+,\beta_-)= 3
\left\{
e^{-8 \beta_+} - 4 e^{-2 \beta_+} \cosh(2 \sqrt{3} \beta_-)
+ 2 e^{4 \beta_+} \left[ \cosh( 4 \sqrt{3} \beta_-) -1 \right]
\right\}.
\end{eqnarray}
{\em In particular, for all Mixmaster solutions whose trajectories
in the anisotropy plane at least once pass through the
origin, $\beta_+ = \beta_-=0$, the constant of motion
$K_{\psi}=0$ exactly.
}

Is $K_\psi$ exactly proportional to the minisuperspace Hamiltonian?
We do not know, but it seems to us that,
 considering the complexity of the calculations involved,
it would be an extraordinary coincidence if the first two terms in
the expansion of $K$ coincided with those of $H$, without $K$ and $H$
being essentially equal (proportional). In addition, if all Mixmaster
trajectories passed through the origin the value of $K_{\psi}$
would have to be exactly
equal to zero for all Mixmaster initial data. Regrettably, it is not
known how large is the set of Mixmaster trajectories passing through
the origin.

Does this indicate that the constant of motion $K$ is proportional to the
hamiltonian constraint, i.e., uninteresting and trivial, for all
$U(1)$-symmetric spacetimes on $S^3$? Is it weakly zero?
Not at all! To see that, we shall
calculate $K$ for the Taub-NUT spacetime, using the additional Killing
field the general Mixmaster spacetime does not have.

The Taub-NUT spacetime is a special case of the Mixmaster spacetime,
obtained by putting $A=B$ in (\ref{ds2}). Its equations of motion
can be explicitly solved,  and give the solution (written here in
the right-invariant form, slightly
different form \cite{hawking-ellis}, where it is given
in the left-invariant form):
\begin{equation}
\begin{array}{ll}
ds^2&= -U^{-1} dt^2 +  (t^2 + l^2)
\left[ (\hat\sigma^1)^2 +(\hat\sigma^2)^2 \right]
+ (2 l)2 U (\hat\sigma^3)^2, \\
&U= \displaystyle \frac{ -t^2 + 2mt + l^2}{ t^2 + l^2}.
\end{array}
\end{equation}

Due to the additional  restriction ($A=B$),  the Taub-NUT spacetime has four
Killing fields; the three left-invariant vector fields inherited from
the Mixmaster plus one of the right-invariant vector fields, namely
the $\partial_\phi$. Note that
$\partial_\phi$, being a right-invariant vector
field, commutes with other three Killing fields.
  This additional Killing field, as mentioned in
the previous section, also generates a
$U(1)$ group action which induces a Hopf fibration of $S^3$. We
can use it (actually $2\;\partial_\phi$ because of the normalization
convention we use)
to calculate a new constant of motion, which we shall call
$K_\phi$. Owing to the simplicity of the Taub-NUT solution
all angular integrals can now be
easily calculated and the constant of motion turns out to be:
\begin{equation}
K_\phi[{\rm Taub}] = 16 (8\pi)^2 (2l)^2 (l^2 + m^2).
\end{equation}

As we can see, $K_\phi$[Taub] is a strictly positive, non-trivial function
of the  initial data ($l$ and $m$) and as such
cannot be proportional to the Hamiltonian constraint, which is zero for
all solutions of the Einstein equations. More generally, it guarantees
that $K_\phi$ is not weakly zero.  This implies that there
exists a neighbourhood $\cal O$
of the Taub-NUT initial data, open in the set of all
$U(1)$-symmetric initial data on $S^3$ with $\partial_\phi$ as a
Killing field, in which the  constant of motion $K_\phi$ is strictly
positive.
That constant of motion is local in time as a functional of the usual
gravitational initial data---the three-metric and its momentum---and
can be written down essentially explicitly. It seems to be the only
explicitly known observable for such a large family of Einstein
spacetimes.

For the left-invariant Mixmaster metric---which,
as opposed to its isometric
right-invariant counterpart (\ref{ds2}),
has $\partial_\phi$ but not
$\partial_\psi$ as Killing field---the $K_\phi$ constant of motion
can be calculated.
(The left-invariant Mixmaster metric can be obtained from
(\ref{ds2}) by making $A$, $B$ and $C$ appropriate functions of the
angles or exchanging $\sigma$-s for $\omega$-s.)
Due to isometry of the left- and right-invariant
spacetimes, it must be that $K_\phi$[Mixmaster]$=K_\psi$[Mixmaster].
Assuming that $K_\psi$[Mixmaster] is indeed proportional to the
Hamiltonian constraint, it follows that the Mixmaster spacetime is the
special case for which the $U(1)$-generated constant of motion is
(weakly) identically zero.

The positivity of $K_\phi$[Taub] also implies that Geroch's transformation
can yield a new globally defined spacetime on $S^2\times S^1$
when applied to Taub-NUT. Since the harmonic variables can be easily
calculated in this case, the transformation can be carried out explicitly
and we find that the new solution on the trivial bundle turns out to
be the Kantowski-Sachs solution.
This had already been established by
Geroch himself \cite{geroch72}, but only locally.

Now a word about the possibility of generating new globally defined
solutions of the vacuum Einstein equations by applying Geroch's
transformation to the Mixmaster spacetime.
Moncrief's analysis of the global
applicability of Geroch's transformation \cite{moncrief87}
guarantees  that it is
possible to transform a solution from $S^3$, as the bundle
with winding number $n=1$, to a new solution on the trivial bundle
$S^2\times S^1$ if $K\geq 0$.
We showed that for all the Mixmaster solutions that pass through
the $\beta_+=\beta_-=0$ point in the anisotropy plane
$K$ is exactly zero; the condition for global applicability of
Geroch's transformation is therefore satisfied and we can obtain a new
solution on the trivial bundle. Likewise, if $K$ is exactly zero for all
Mixmaster spacetimes, as suggested by our perturbative result, a new
solution globally defined on the trivial bundle can be obtained
from any Mixmaster solution.

 What properties does  that new solution have? It is not a
homogeneous solution, for the following reason.
If it were homogeneous it would have to be the
Kantowski-Sachs solution, which is the only homogeneous solution on
$S^2\times S^1$. But the Kantowski-Sachs constant of motion $K$
associated with the Killing vector field tangent to the $S^1$ factor is
strictly positive. It is positive because it must be equal to the
Taub-NUT constant $K$, since $K$ is invariant under the
action of Geroch's transformation and the Kantowski-Sachs solution
is just the transformed Taub-NUT solution.
The same invariance requires that the Kantowski-Sachs $K$, a
strictly positive quantity,  be
equal to the Mixmaster constant $K$, which is identically zero,
leading to a contradiction. So, the new solution on the trivial bundle
is not homogeneous.

The asymptotic behavior of the new solution requires additional
analysis; for now let us just say that it is probably oscillatory in
the BKL sense. The BKL condition for the existence of the oscillatory
behavior is an ``open" condition, i.e., some quantity has to be
different from zero. It is satisfied by the Mixmaster solution, hence
it has to be satisfied by the transformed solution obtained by using
an element of $SL(2,R)$ sufficiently close to unity. Whether the
element of $SL(2,R)$ that gives the new {\em globally} defined
solution on the trivial bundle violates the BKL condition or not, will be
the subject of future study.

\section{Conclusions}

The constant of motion $K$ we calculated in the previous section is
a second degree polynomial in the minisuperspace momenta
$(p_\Omega,p_+,p_-)$ and, to the first order in the Misner anisotropy
variables $(\beta_+,\beta_-)$, it is proportional to the minisuperspace
Hamiltonian. We believe that it is reasonable to conjecture that $K$
is exactly proportional to the minisuperspace Hamiltonian. The
coincidental agreement of the first two terms in the expansion of
$K$ and $H$, considering the complexity of the expression for the
constant of motion $K$, is highly unlikely unless $K$ and $H$ are exactly
proportional. It seems that  the Mixmaster dynamical system has once more
defied the attempt to find a non-trivial constant of motion that would
help us to gain more detailed knowledge of its behavior. Of course,
this might just
be a consequence of the simple, but not yet established,
fact  that there are no non-trivial
constants of motion and that a generic Mixmaster trajectory fills
densely an open subset of the constraint surface in the minisuperspace.
Indeed, Kucha\v{r} \cite{kuchar81} showed that for non-homogeneous spacetimes
there are no (weakly) non-zero constants of motion that are linear
functionals of the gravitational momenta.
Presently K. Schleich and D. Witt \cite{schleich}
are trying to prove that
any constant of motion for the Mixmaster dynamical system  that is
quadratic in minisuperspace
momenta---which includes our $K$---has to be
proportional to the minisuperspace
Hamiltonian, i.e., has to be trivial. Our perturbative result for $K$
supports that.

More importantly,
the probable weak vanishing of $K$ for the Mixmaster spacetime
does not mean that $K$ is weakly zero for all $U(1)$-symmetric
spacetimes.  By calculating explicitly the constant of motion $K_\phi$
for the Taub-NUT spacetime, we showed that around the Taub-NUT intial
data, there exists an open set in the set of all $U(1)$-symmetric
initial data for which the constant of motion
$K$ is a non-trivial spatially non-local, but temporally local, strictly
positive functional of ADM canonical data.
That constant of motion could be used as an
observable in a $U(1)$-symmetric toy model for the
future quantum theory of gravity. Classically, the
value of this constant of motion for a particular $U(1)$-symmetric
spacetime might signal
whether the spacetime is asymptotically velocity-dominated or not.
The value
of K for the Taub-NUT spacetime is positive and the spacetime is
velocity-dominated, on the other hand the value of $K$ for the
Mixmaster spacetime, which is not a velocity-dominated spacetime,
is presumably zero, which might signal that the $K=0$ surface in the
set of $U(1)$-symmetric initial data is the boundary between the
simple velocity dominated behavior and the much more complicated
oscillatory BKL-like behavior.  The Mixmaster spacetime would then
be exactly at the boundary of the set of velocity-dominated spacetimes.
Likewise, an interesting and closely related issue we would like to resolve
is: Does Geroch's transformation, which does not change the value of
$K$, ``conserve'' the velocity-dominated behavior. i.e., are new
solutions obtained from velocity-dominated  solutions also
velocity-dominated?

\acknowledgments
Both authors are grateful to the Institute for Theoretical Physics in
Santa Barbara, where part of this research was carried out, for its
support and hospitality.
This work was supported in part by NSF Grants
PHY-9201196 to Yale University and PHY-890435 to Institute for
Theoretical Physics.


%
%
\end{document}